# A high -quality and -throughput colloidal lithography by mechanical assembly and ice-based transfer

Sivan Tzadka[1,2], Abed Al Kader Yassin[3], Esti Toledo[1,2], Jatin Jawhir Pandit[1,2], Angel Porgador[3], Mark Schvartzman[1,2*]

[1]Department of Materials Engineering, [2]Ilse Katz Institute for Nanoscale Science and Technology, Ben-Gurion University of the Negev, P.O. Box 653, Beer-Sheva 84105, Israel; [3]The Shraga Segal Department of Microbiology, Immunology, and Genetics Faculty of Health Sciences, Ben-Gurion University of the Negev, P.O. Box 653, Beer-Sheva 84105, Israel

* Corresponding author (marksc@bgu.ac.il)

**Abstract**

Colloidal lithography has emerged as a promising alternative to conventional nanofabrication techniques, offering the ability to create nanoscale patterns in a cost-effective and scalable manner. However, it has been so far limited by defects such as empty areas or multilayered regions, hindering its application. We introduce a novel "ice-assisted transfer" technique that combines rubbing-based particle assembly on elastomer substrates with ice-mediated transfer to achieve defect-free, high-quality polycrystalline particle monolayers. This approach eliminates foreign material contamination and enables precise control of particle arrangement and density. By optimizing process parameters, including surfactant concentration and water film thickness, we minimized defects and demonstrated the versatility of this method in fabricating functional nanoscale structures. We highlighted the benefits of this process through two applications: (1) antireflective "moth-eye" coatings, which achieved near-zero reflection in the mid-infrared spectrum due to improved particle monolayer quality; and (2) nanostructured surfaces for ligand-free T-cell activation, whose topography enhanced cell activation, showcasing potential for immunotherapy applications. The process achieves rapid, cost-efficient patterning without requiring specialized equipment, making it suitable for diverse fields requiring scalable nanostructuring. This work represents a significant advancement in colloidal lithography, addressing critical challenges and unlocking its potential for practical applications in optics, biotechnology, and beyond.

Over the last two decades, colloidal lithography has been recognized as an attractive alternative to traditional bottom-up nanofabrication methods due to its ability to produce quasiperiodic nanoscale patterns in a simple, low-cost manner, without the need for sophisticated equipment[1,2]. This assembly can be performed using various self-assembly approaches[3,4], such as dip-coating[5], spin-coating[6], solvent evaporation[7], and Langmuir-Blodgett based assembly[8], and broadly applied in engineering of nanoscale devices in systems in various areas including but not limited to photonics[9], photovoltaics[10], plasmonics[11], nanofludics[12], and sensing[13]. However, despite the extensive research in the field of particle self-assembly, it has not yet been utilized for scalable fabrication of nanoscale devices and systems, beyond research prototypes. The main reason for this is its inability to produce defect-free structures over large areas. While self-assembly can produce perfectly ordered architectures in short-range order, long-range order is plagued by defects such as empty patches or regions with multilayered particles. These defects render the assembled structures unsuitable for practical applications. For example, while colloidal lithography holds promise for optical applications, such as antireflective moth-eye structures[14,15], structural defects act as scattering centers or diffraction gratings, preventing these structures from achieving their optimal antireflective performance as predicted by theory[16,17]. Similarly, defects produce diffusion scattering and degrade optical quality in self-assembled diffraction gratings and photonic crystals[9].

It is important to note that conventional colloidal lithography methods are based on particle self-assembly from the liquid-air interface. Alternatively, micro- and nanoparticles can be assembled using a "dry approach," which involves mechanical rubbing between two surfaces. This method organizes particles into ordered arrays driven by particle-particle and particle-surface attraction forces. Early demonstrations of rubbing on solid surfaces yielded limited particle organization[18,19]. However, later advancements using two elastomeric (PDMS) substrates achieved polycrystalline monolayers of Polystyrene with unprecedented quality, surpassing what is possible through wet-phase assembly[20]. Since then, there has been a growing interest in rubbing-based particle assembly, with recent works expanding this approach to particles of non-spherical shape[21], particle material such as silica and PMMA[22], and applications in various areas including photonics[23], biology[24], and sensing[25]. Despite these advancements, rubbing-based methods of nano/micro- particle assembly into dense periodic array are still limited mostly to PDMS substrates, significantly restricting the potential applications of this promising fabrication method. An alternative approach includes assembly of Silica particle on other substrates coated with molecular glue such as Polyethyleneimine (PEI), with subsequent PEI pyrolysis in high temperature, but this process is incompatible with polymer particles[18]. To circumvent this limitation, we recently demonstrated that particle monolayers assembled on PDMS could be mechanically transferred to other substrates using thin PEI which can be then removed by dissolution[26]. However, this process is highly sensitive to glue thickness and process conditions, with the residual PEI layer severely limiting the pattern transfer through the particle mask and causing undesirable contamination when attempts were made to remove it. Thus, there is a critical need for a mechanical assembly and transfer method that can produce clean, defect-free particle assembly without introducing foreign materials. Such a method would unlock the potential of this fabrication approach for broader and more practical applications.

Here, we introduce a novel colloidal lithography approach for fabricating micro- and nanoscale functional structures with highly controlled organization and density, which we term **"ice-assisted transfer"** (schematically illustrated in Fig. 1a). The process begins with the dry rubbing of nano- or microparticles between two PDMS surfaces, resulting in both surfaces being coated with highly dense and ordered polycrystalline particle monolayers. Subsequently, a drop of water is placed onto the monolayer-coated PDMS surface, and the target substrate is placed above and gently pressed against the PDMS substrate, resulting in the formation of a thin uniform water film sandwiched between the two surfaces. This "sandwich" is then frozen below the melting point of water, embedding the particle monolayer in the ice layer. The PDMS is then peeled off, leaving the ice film with embedded particle monolayer attached to the target surface. Finally, the ice is allowed to melt at room temperature, followed by gradual water evaporation, producing a high-quality, high-density polycrystalline particle monolayer on the target substrate. This process is performed without introducing any additional materials, ensuring a clean transfer process that does not interfere with subsequent patterning or fabrication steps. We first optimized the process parameters, and mostly the addition of surfactant to water, to obtain large area monolayers of particles of different sizes with high quality and lowest possible defect density (Fig. 1b). We then demonstrated the efficiency and versatility of this fabrication approach in two key applications: antireflective sub-wavelength structures for precision optics, and 3D structures designed for the mechanical activation of immune cells, which hold significant potential for advancing immunotherapy. Overall, this fabrication method unleash the potential of bottom-up nanostructuring for the development of engineered advanced nanoscale materials and functional systems.

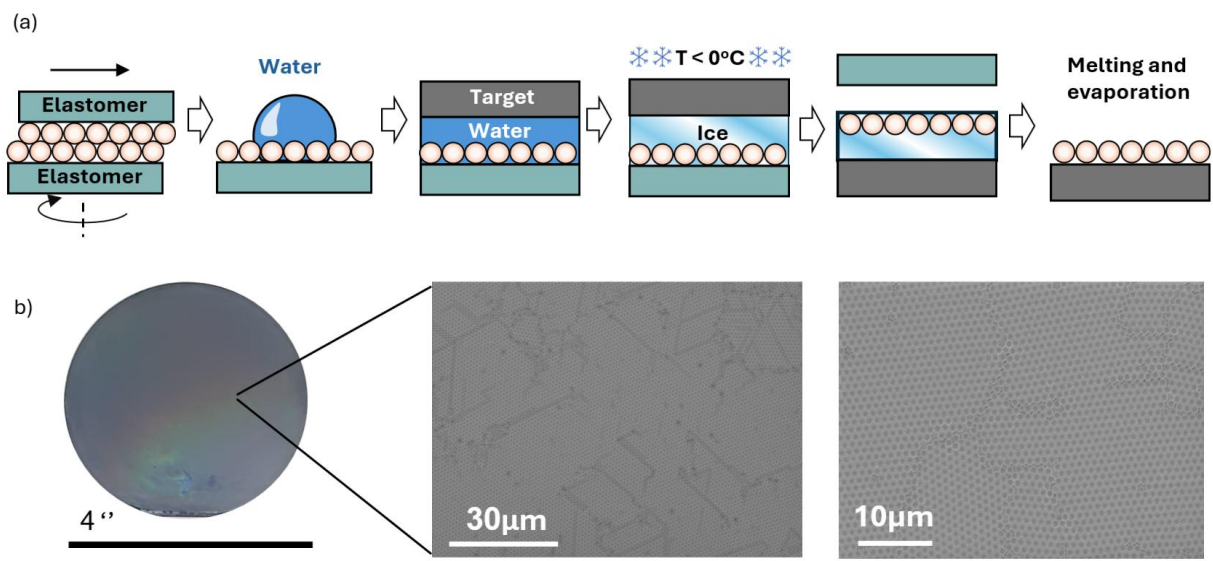

Figure1. (a) Process flow describing the rubbing-based assembly and pattern transfer using ice film. (b) Image of a 4 inch Silicon wafer covered with ice-transferred particle monolayer and SEM images of the monolayer.

To produce monolayers, we used polystyrene nanoparticles of various sizes, ranging from 200 nm to 2 µm. We found that the quality and robustness of the process depend on several parameters, the most critical being the amount of water used for the transfer, and the addition of surfactant. Through empirical observation, we established a rule of thumb for the optimal water amount: the water film thickness should be approximately *20d+10 µm*, where *d* is the particle diameter in microns. The thickness, in turn, can be precisely controlled through the volume of the water drop, which can be calculated considering the area of the target-PDSM interface. In all the experiments described herein, the natural weight of the target substate was enough to flatten the drop to an uniform film. Also, gentle Oxygen plasma treatment, which was usually used to clean the target surfaces prior to the sandwiching, produces hydrophilic surfaces that can facilitate the uniform water spreading. However, we found that uniform water film is formed also without such plasma treatment, probably due to mechanical weight of the target. The obtained water film that is sufficiently thick to embed the particles but thin enough to allow for reasonable drying times. The final step of the process is technically comparable to self-assembly from a liquid phase, similar to conventional methods such as dip-coating or Langmuir-Blodgett assembly. In this step, particle organization is driven by a well described mechanism highly dependent on particle-particle, particle-liquid, and particle-substrate interactions[27], which are influenced by particle charge and wettability.

The zeta potential of the nanoparticles, measured using dynamic light scattering, was approximately −30 mV (exact values are provided in Table S1), imparting a negative charge to the particles. This negative charge causes the particles to be attracted to the water solution during drying, leading them to flow with the water rather than maintaining a monolayer organization as formed during rubbing. To counteract this effect, the pH of the water-based solution was adjusted to 10 by adding ammonium hydroxide, which imparted a negative charge to the solution, repelling the negatively charged particles. Additionally, particle wettability was modified by adding a nonionic surfactant based on Octyl Phenol Ethoxylate (Triton TX-100) at varying concentrations.

The effect of surfactant concentration qualitatively mirrors its impact as reported for assembly from the liquid phase[28]. Without surfactant, the resulting monolayer exhibited frequent and relatively large micron-scale defects in the form of empty patches (Fig. 2a). Adding surfactant at a concentration of 0.075 mM significantly reduced both the size and frequency of these defects. A further increase to 0.25 mM produced monolayers with minimal defects and large, well-defined monocrystalline domains. This structure is comparable to those formed in the initial stages of rubbing-based assembly on PDMS (Fig. S1). However, increasing the surfactant concentration to 0.4 mM led to the formation of particle clusters on top of the monolayer (double-layer domains), which in turn created vacancies within the monolayer. Such clusters have been reported for the nanoparticle assembly from liquid phase [29,30]. One proposed mechanism of their formation is related to the release of stresses in the nanoparticle lattice, which are formed due to simultaneous formation of the lattice domains on the solid surface at different places and with different orientations[31]. An alternatively proposed mechanism is associated with the accumulation of the particles on top of interstices formed in the first-layer lattice as the result of occasional sinking of some particles prior the liquid drying[32]. Notably, while Fig. 2a illustrates monolayers formed using 1 μm particles, similar qualitative trends in monolayer quality as a function of surfactant concentration were observed for other particle sizes as well, such as 500 nm particles shown in Fig. S2.

To quantify the quality of sphere arrangement as a function of surfactant concentration, we analyzed the SEM images using Fast Fourier Transform (FFT), as described in detail in SI. For patterns obtained from pure water, as well as from water with 0.075 mM and 0.4 mM Triton concentrations, the FFT analysis showed ring patterns characteristic of polycrystalline structures with randomly oriented domains (Fig. 2a - insets). In contrast, the optimized concentration of 0.25 mM produced an FFT pattern indicative of uniform crystalline orientation, demonstrating superior monolayer quality. To further validate this, we quantified the FFT results by measuring the width of the first-order FFT peaks (Fig. 2b). The comparison of peak broadening confirmed that 0.25 mM surfactant concentration produced the highest order and quality of the formed particle monolayer.

Also, the monolayer quality was quantified using Voronoi tessellation[33] (Fig. 2c). This method involves determining the centers of the assembled particles as seed points, which divide the space into polygons, with each polygon containing all points closer to its seed point than to any other. This analysis allows for clear visualization of particle symmetry and defect types. The order of the assembly was further quantified using the **Voronoi Regularity Index (RI)**, defined as the ratio of the standard deviation of polygon areas to their average area, and the **Areal Disorder (AD) factor**, calculated as:

$$AD = 1 - \left(1 + \frac{\sigma}{\bar{A}}\right)^{-1} \tag{1}$$

where $\bar{A}$ and $\sigma$ are the average and standard deviation of polygon areas, respectively, with *AD=0* indicating perfect particle order. Both the regularity index and the areal disorder factor were analyzed using at least six images for each Triton concentration and were found to reach their minimum values, as expected, at a concentration of 0.25 mM (Fig. 2e and f). Also, we analyzed the effect of surfactant concentration on defect occurrence (Fig. 2g). To differentiate defects caused by missing particles from empty areas at naturally occurring grain boundaries, defects were defined as empty regions with a minor axis larger than the sphere diameter. Figure 2d shows the total defect area as a function of Triton concentration, which also minimized at 0.25 mM.

Finally, the average size of the grains formed in polycrystalline assembled structures is an additional merit of the assembly quality and order. Here, the largest grain size was obtained, as anticipated, for 0.25 mM surfactant concentration (Fig S3.) in Note, the bare polystyrene particles, lacking any functional groups covering their surface, were used in this study. The possible effect of such groups on the first part of the process (i.e., the rubbing step) is unclear and will be the subject of further investigation. As for the second part of the process, which involves particle monolayer formation during water evaporation, the water content—specifically, pH and surfactant concentration—may require further optimization depending on the functional groups present, as these groups alter the particle's zeta potential and influence their tendency to aggregate[34].

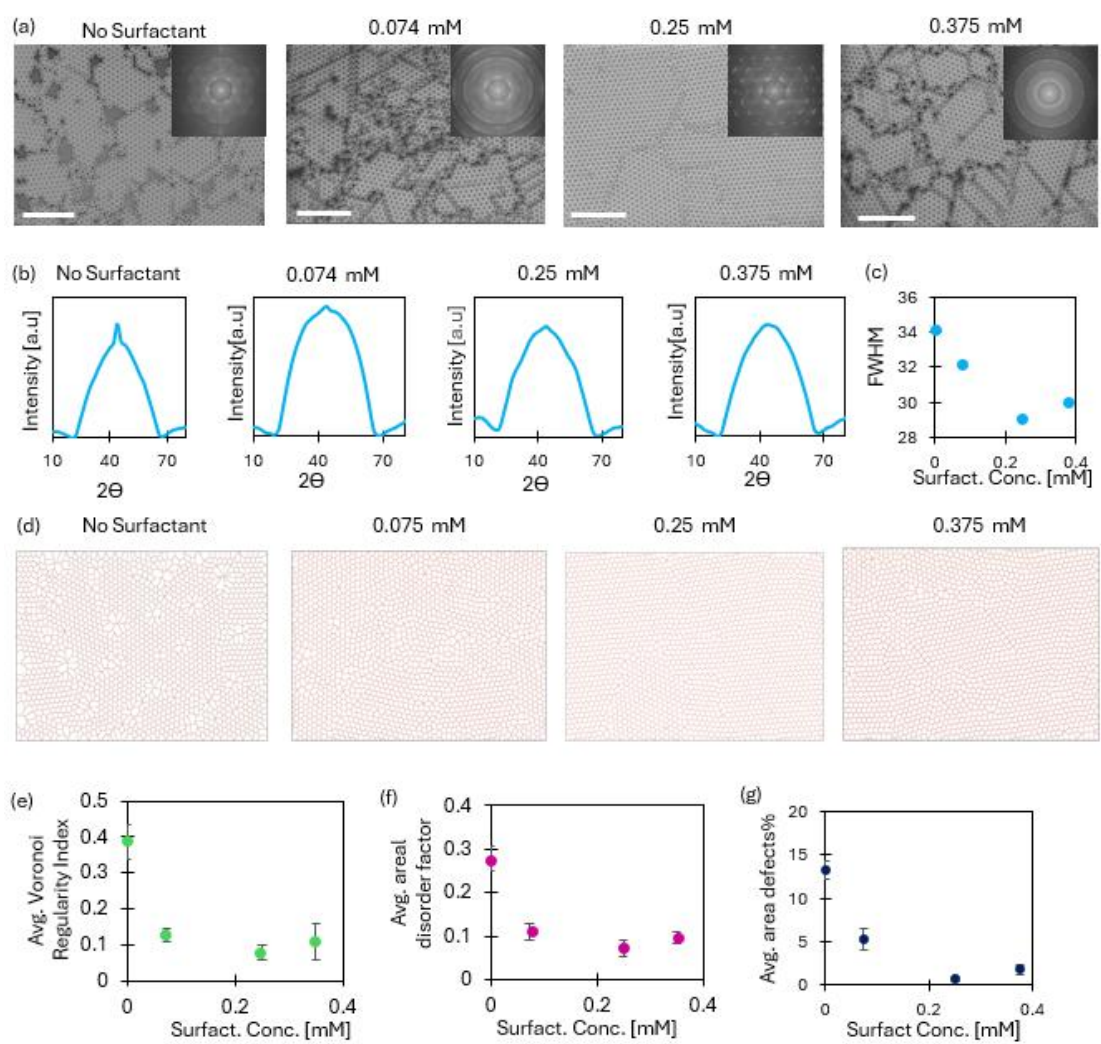

Figure 2: **Effect of Surfactant on the quality of transferred particle monolayer.** (a) SEM image of monolayers obtained for various surfactant concentrations. Insets show FFT patterns for each case. Scale bar: 10 microns. (b) Width of the first order FFT peaks, as a merit for the monolayer uniformity, for each surfactant concentration. (c) Width of the first order FFT peak vs. surfactant concentration. (d) Voronoi tessellations of the monolayers shown in (a). (e) Voronoi regularity index vs. surfactant concentration. (f) Areal disorder factor vs. surfactant concentration. (g) % of area occupied by defects vs. surfactant concentration. The data shown in (d)-(f) were averaged for 10 SEM images taken for each surfactant concentration.

It should be noted that the demonstrated nanofabrication approach produces high-quality patterns within an impressively short processing time (a few minutes), in a cost-effective manner, and without requiring specialized equipment. Furthermore, this approach has many advantages over the previously demonstrated fabrication methods based on rubbing-based colloidal assembly, including recently published by us colloidal assembly combined with PEI-based transfer PEI[26]. In particular, PEI layer connecting the micro-/nano-particles to the target surface often produces an obstacle for the pattern transfer method. Fig. S4 shows SEM images of Silicon etched though colloidal mask produced by PEI-assisted transfer, in which the pattern distortion due to PEI residuals are clearly seen. Similarly, PEI layer produces an obstacle for the pattern transfer by thin film deposition and liftoff. In functional application of the transferred colloids, PEI (or any other molecular glue) can attract to undesirable particle contamination.

Notable, PEI can be removed after the transfer by dissolution in ethyl alcohol, but this process is long (up to a few hours), barely controlled, and can lead to partial detachment of colloidal particles from the surface. On the contrary, ice assisted transfer method produces clean interface between the particle and target surface, with no presence of foreign material, ensuring easy and straight forward pattern transfer by any existing nanofabrication approach. The overall process is quick, and mostly depends on the freezing time, which takes about 15 – 30 minutes in standard freezer (-4°C), but can be shortened to up to 5 minutes by using -80°C freezer. Also, the melting and evaporation stage of the process can be expedited to 2-3 minutes by placing the surface to vacuum environment, such as pumped desiccator.

These advantages of the ice-assisted colloidal transfer makes it well-suited for numerous applications requiring scalable nanopatterning, especially those incompatible with the expensive technologies used in microchip production. Here, we demonstrate two such applications. The first is the fabrication of antireflective nanostructures, commonly referred to as "moth-eye" structures due to their bioinspiration from the cornea of moths[35]. These structures consist of dense arrays of sub-wavelength cones or nipples. When scaled significantly below the wavelength of light, these structures act as an effectively homogeneous optical medium, with a refractive index gradient determined by the shape of the structures[36]. Properly designing the structure profile can achieve a broadband and omnidirectional antireflective effect that is unachievable with standard thin-film-based antireflective coatings. Additionally, moth-eye structures offer superior durability under thermomechanical stresses compared to antireflective films. Despite extensive research in this area, practical realization has been hindered by the lack of cost-effective yet reliable methods to fabricate these structures with controlled size and distribution.

In this study, we designed moth-eye structures on silicon for the mid-IR spectrum, targeting a minimum reflectance around 2.8 microns. To achieve this, we assembled 1 µm particles on silicon using the described method with a Triton concentration of 0.25 mM. Particles assembled using pure water served as a control to demonstrate the importance of appropriate surfactant concentration for achieving optimal application outcomes. The particle monolayer was transferred to silicon, and used as a mask for silicon dry etching, which was followed by particle removal with acetone (details of all the fabrication processes appear in SI). SEM images of the etched moth-eye structures produced form colloidal masks assembled with and without Triton are shown in Fig. 3a and 3b. Fig. 3c presents high-magnification SEM images of the etched structures. Importantly, the structures produced with and without the surfactant were fabricated using the same etching process and exhibited identical geometry. Notably, the little elevated discs on top of the pillars are the result of the plasma under-etch in the Silicon areas which were not in the complete contact with the masking polystyrene spheres.

The obtained moth-eye structures were characterized by measuring the specular reflectance spectrum in the range of 2.5 to 3.5 microns (Fig. 3d). In both cases, substantially reduced reflection was achieved compared to bare silicon. Interestingly, the differences in the monolayers formed with and without surfactants were directly reflected in the antireflective performance of the fabricated structures. The surface structures produced using the surfactant-assisted method exhibited an almost zero-reflection spectrum, with a minimum reflectance at approximately 2.8 microns. This minimum closely matched the theoretical reflection spectrum simulated using the transfer matrix method[37]. In contrast, identically shaped antireflective structures fabricated using pure water during the transfer process showed a minimum reflectance of around 5%. This imperfect antireflection effect is attributed to excessive defects in the microsphere monolayers produced without surfactants, particularly in the form of relatively large patches with missing particles. These defects were transferred to the etched patterns, creating unpatterned areas that contributed to overall reflection. Additionally, the numerous defects scattered light, further increasing the measured reflection[17,38]. These results clearly demonstrate the importance of using an appropriate surfactant in the sphere transfer process: the surfactant not only improves the visible quality of the resulting microstructure but also significantly enhances its functional performance.

We also applied this mechanical assembly method to produce broadband and omnidirectional antireflective structures on sapphire for the mid-IR spectrum. In this case, the structures were designed as periodic holes with a conical cross-section. The fabrication process involved sphere transfer onto sapphire coated with a Poly-(butyl methacrylate) (PBMA) sacrificial film, followed by sphere diameter reduction using oxygen plasma, nickel evaporation (~40 nm), and the subsequent lift-off of PBMA and spheres. Here, the sacrificial PBMA layer (~100 nm) was added to produce an undercut in the nanosphere monolayer, and facilitate liftoff after the evaporation of the relatively thick metal film. The sapphire was then plasma-etched through the nickel mask, which was then stripped to reveal the final structure (Fig. S5a). Figure 3e shows the etched structures, featuring a conical shape with a depth of 600 nm (Fig. S5 b-c). Figure 3f presents the reflectance spectrum of the sapphire surface patterned with the antireflective structures, alongside the simulated spectrum of these structures (using the same method as used for silicon) and the spectrum of bare sapphire. The pronounced broadband antireflective effect observed in the patterned surface is likely due to the smooth refractive index gradient created by the sidewall profile of the etched holes, as also confirmed by the simulation.

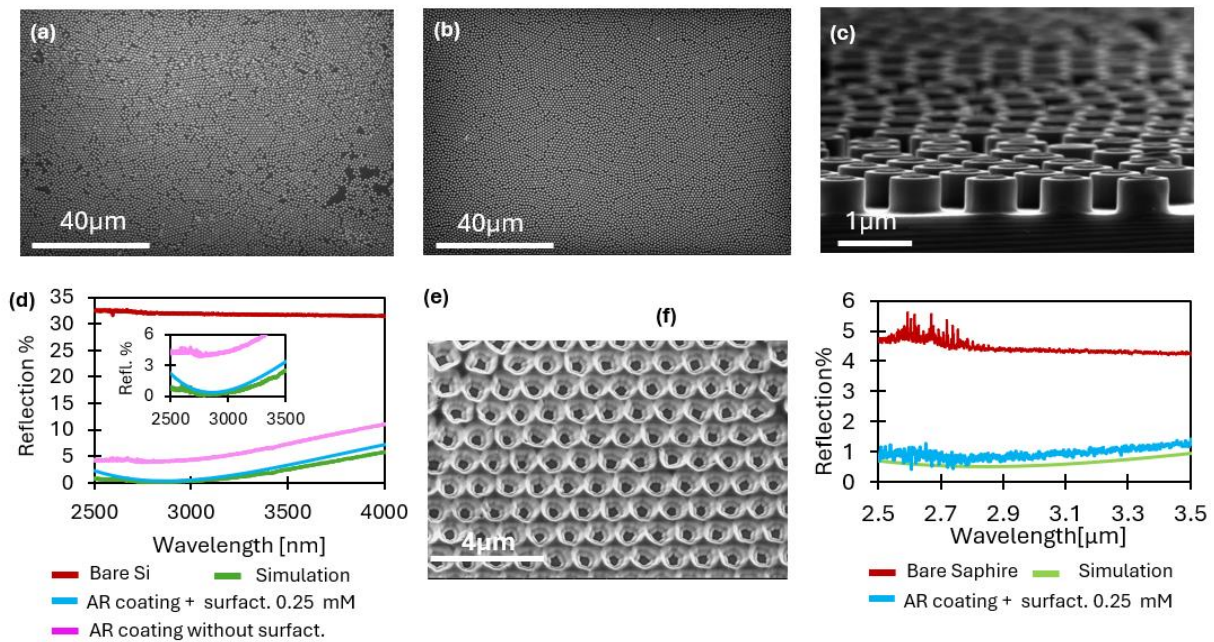

**Figure 3. Applications of the ice-assisted transfer method in antireflective structures** (a) Antireflective structures on Silicon produced without surfactant, showing defects that were transferred from the colloidal mask to Silicon by dry etching. (b) Similar structures as in (a) produced using surfactant (0.25 mM), with almost no defects. (c) Close – up SEM of the antireflective structures, showing their side wall profile and dimensions. (d) Optical characterization of the antireflective structures, with the effect of the surfactant addition on the obtained reflection attributed to the reduction in scattering defects. (e) Broadband antireflective structures on Sapphire (f) Optical characterization of the antireflective structures on Sapphire.

The second application showcasing the benefits of our novel nanofabrication approach lies in the realm of nanostructured materials designed to interact with biological systems. T cells, a vital component of the adaptive immune system, play a crucial role in defending the body against pathogens, viruses, and stressed or mutated cells. **In vivo**, T cells are activated through the specific binding of T-cell receptors (TCRs) and costimulatory receptors to ligands expressed on antigen-presenting or target cells. Beyond these chemical ligand-receptor interactions, the physical characteristics of T-cell-target interfaces, such as the mechanical stiffness of activating surfaces and the nanoscale topography of the interacting membranes, significantly influence T-cell activation[39–41]. While the precise mechanisms underlying these physical effects remain unclear and are under intensive study, it is suggested that nanoscale protrusions on the T-cell membrane (microvilli) facilitate the separation of TCRs, concentrated at the protrusion tips, from large phosphatase molecules that inhibit activation signaling[42–45].

Surfaces structured with nanoscale topography and coated with activating ligands were shown to induce cell protrusion and enhance activation of T cells and of other lymphocytes such as natural killer (NK) cells[46–50]. Interestingly, the nanoscale topography of activating surfaces not only enhances ligand-induced activation but can also independently activate T cells to some extent, even in the absence of activating ligands[51]. While the mechanism behind such "purely mechanical" activation is not yet fully understood, it is evident that nanoscale topography can and should be harnessed for designing material-based platforms for *ex-vivo* T-cell activation. These platforms could be used both for mimicking T-cell environments in fundamental studies of activation mechanisms and for producing immunotherapeutic T cells with enhanced antitumor activity.

A major challenge in employing nanotopography for *ex-vivo* T-cell activation platforms lies in their scalable fabrication. Top-down nanofabrication approaches adapted from electronic device production are extremely expensive and slow, making them unsuitable for producing large-area surfaces required for clinical-scale cell activation. We believe that our novel fabrication approach can address this challenge. To demonstrate this, we created large-area surfaces structured with arrays of nanowells designed to facilitate the formation of T-cell protrusions (Fig. 4a). This involved forming monolayers of 1 µm polystyrene nanospheres, reducing their diameter controllably using oxygen plasma, etching silicon through the resulting sphere masks, and stripping the nanospheres to create arrays of silicon nanopillars. These pillars were then negatively replicated onto polydimethylsiloxane (PDMS) to produce PDMS surfaces patterned with arrays of pores. The pore sizes were determined by the initial pillar diameter, and the periodicity was controlled by the original sphere diameter. These patterned surfaces were used for ligand-free T-cell stimulation. Pores with diameters of 200 nm, 400 nm, and 550 nm were fabricated, each with a periodicity of 1 µm (Fig. 4b and Fig. S6). Flat PDMS was used as a control to provide an environment devoid of topographic cues.

We isolated peripheral mononuclear blood cells from a healthy donor and incubated them on the fabricated surfaces for 24 hours. The cells were then fixed, stained for membrane, cytoskeleton, and nucleus, and imaged using z-stack confocal microscopy (Fig. 4 c-e). All topographically patterned surfaces promoted the formation of T-cell nanoprotrusions that penetrated the pores, as clearly seen in the z-stack cross-sections. In parallel experiments, the cells were removed after incubation, labeled with antibodies against CD3 (a T-cell marker) and CD69 (an activation marker), and analyzed using flow cytometry to detect CD69 expression among CD3+ cells. From this data, we calculated the percentage of CD69-positive cells (detailed gating strategy in SI), shown in Fig. 4d.

Flat PDMS produced a negligible amount of about 2% CD69 positive cells, likely due to tonic signaling of T cells. Pores with diameters of 400 nm and 550 nm also produced negligible activation, similar to flat PDMS. However, pores with a diameter of 200 nm increased the proportion of activated cells four folds. While the overall activation level was still lower than that typically achieved using ligand-coated surfaces that trigger activating and costimulatory receptors, the topography-induced activation effect is striking. Furthermore, the observed activation was size-dependent, with only certain pore sizes (200 nm) effectively inducing activation. This phenomenon mirrors findings by Aramesh et al., who reported similar size-dependent ligand-free activation using anodized alumina surfaces[51]. The mechanism behind pore-size-dependent ligand-free activation remains to be elucidated. It has been suggested that pores with diameters of 200 nm or smaller facilitate segregation between TCRs, concentrated at the protrusion tips, and large phosphatases such as CD45, due to the size of the latter[51]. This segregation, in turn, is believed to be the key factor in the mechanism of T cell activation[52]. We believe that the demonstrated fabrication route provides precise control over topographical geometry, enabling fine-tuning of topography-induced activation. This capability could pave the way for designing scalable platforms for fundamental research and clinical applications in immunotherapy.

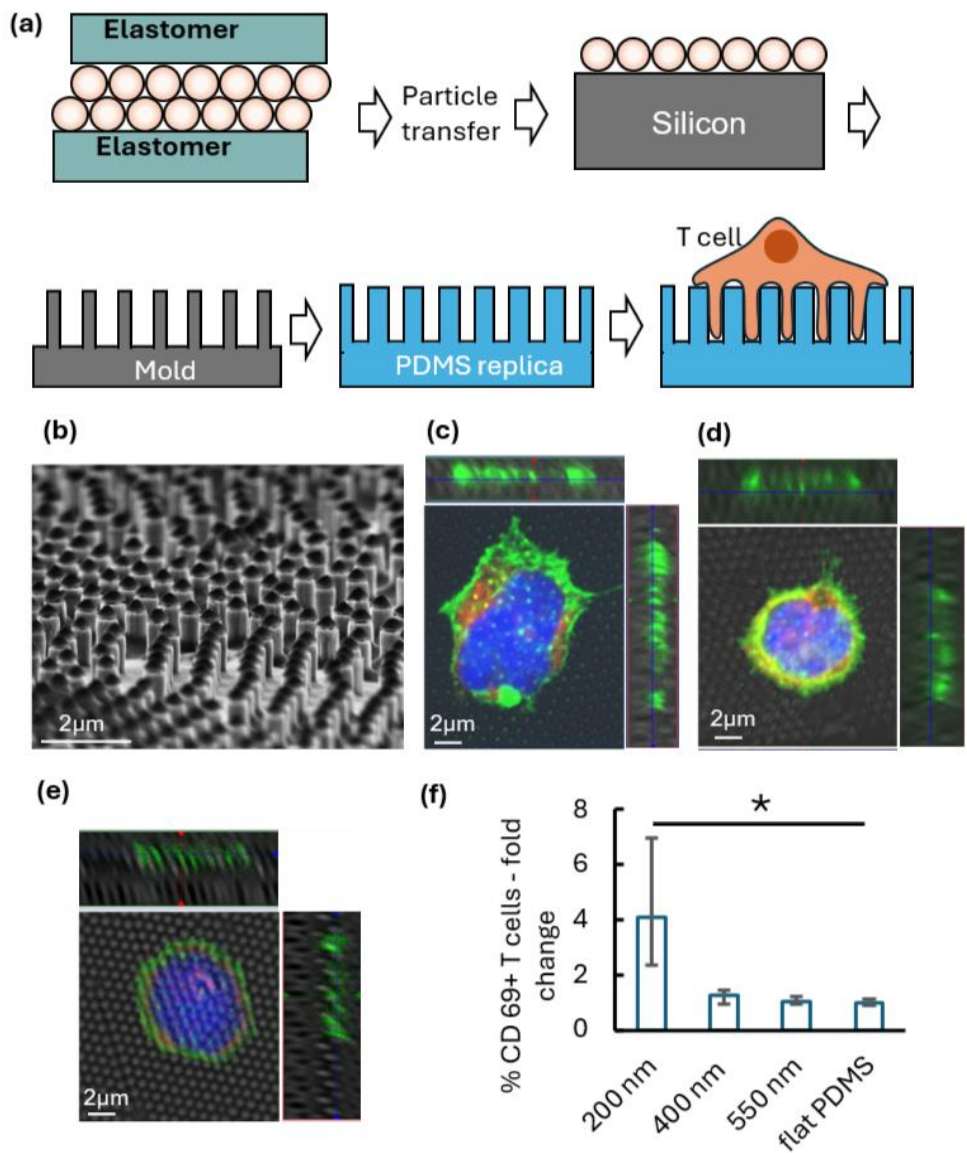

Figure 4. **Nanotopography for T cell activation:** (a) Process flow: fabrication of the structures by PDMS replication. (b) SEM of the mold containing arrays of pillars of 400 nm diameter. (c) – (e) z stack confocal microscope of T cells stimulated on arrays of pores with the diameter of 200 nm, 400 nm, and 550 nm, respectively (f) Effect of the pore size on T cell activation, expressed as the fold change in % of activated (namely CD69+ positive) T cells relative

to the flat PDMS. The data was obtained in triplicates, and analysis was performed with Tukey's multiple-comparison tests using the GraphPad Prism software. *$p < 0.05$

In summary, the fabrication methods described in this paper offer significant advantages in producing high-quality, scalable, and cost-effective nanoscale patterns without the need for specialized equipment. By employing the novel "ice-assisted transfer" technique, the approach ensures clean, defect-free transfer of particle monolayers onto target substrates, overcoming the limitations of traditional colloidal lithography methods, which often produce defects over large areas. The process also allows precise control over particle arrangement and defect minimization through optimized surfactant-assisted assembly. This versatility enables the fabrication of functional structures, two examples of which were demonstrated here: antireflective "moth-eye" coatings for precision optics and nanostructured platforms for T-cell activation, demonstrating its utility across diverse applications. Additionally, the process achieves rapid patterning within minutes. This timescale is much shorter than that of state-of-the-art colloidal lithography techniques, such as those based on Langmuir-Blodgett or dip-coating methods and is more comparable to the timescale of high-throughput lithography. On the other hand, the demonstrated process provides a patterning resolution (i.e., minimal separation between patterned objects) down to the sub-100 nm scale. This level of resolution is unachievable by standard UV-based photolithography and can only be reached using advanced deep-UV or extreme-UV photolithography. However, the latter are extremely expensive in terms of both equipment and operation, making them incompatible with applications beyond microchip production—such as the micro-/nanopatterning for optics or biomedicine demonstrated here. Overall, the unique combination of simplicity, low cost, high pattern resolution, excellent pattern quality, and fast processing time makes this method ideal for applications requiring scalable nanopatterning while remaining cost-efficient and adaptable for practical implementations across a wide range of fields requiring material shaping at the nanometric scale.

# A high-quality and -throughput colloidal lithography by mechanical assembly and ice-based transfer – Supporting Information

Sivan Tzadka[1,2], Abed Al Kader Yassin[3], Esti Toledo[1,2], Jatin Jawhir Pandit[1,2], Angel Porgador[3], Mark Schvartzman[1,2*]

[1]Department of Materials Engineering, Ben-Gurion University of the Negev, P.O. Box 653, Beer-Sheva 84105, Israel; [2]Ilse Katz Institute for Nanoscale Science and Technology, Ben-Gurion University of the Negev, Beer-Sheva, Israel

[3]The Shraga Segal Department of Microbiology, Immunology, and Genetics Faculty of Health Sciences, Ben-Gurion University of the Negev, P.O. Box 653, Beer-Sheva 84105, Israel

* Corresponding author (marksc@bgu.ac.il)

Table S1. Measured Zeta-potential of differently sized PS nanoparticles

Table S1. Zeta potential of nanoparticles

| Diameter(um) | Zeta potential(mv) |
|---|---|
| 0.2 | -30 +/- 9 |
| 0.5 | -27 +/- 6 |
| 1 | -26 +/-3 |
| 2 | -32 +/-5 |

### (a) 1 µm PS particles rubbed on PDMS

Avg. areal disorder factor: 0.15    Voronoi Regularity Index: 0.18    FWHM=28

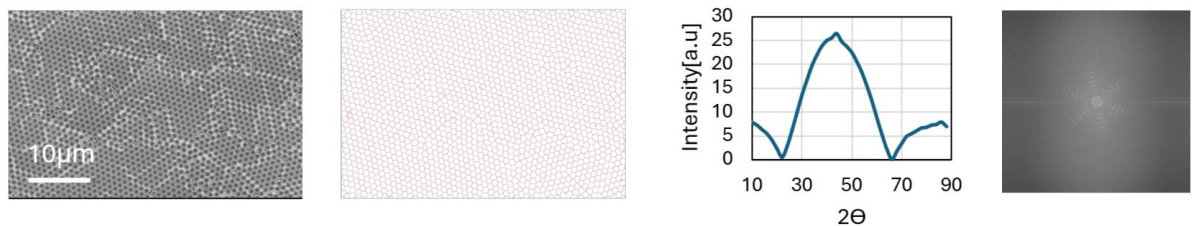

### (b) 500 nm PS particles rubbed on PDMS

Avg. areal disorder factor: 0.17    Voronoi Regularity Index: 0.18    FWHM=31

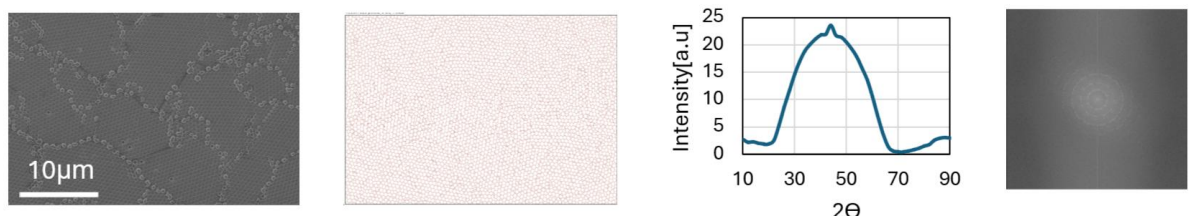

Figure S1. Representative SEM image PS particles rubbed on PDMS, Voronoi tessellation, and FFT of the SEM including the analysis of the FWHM of its peaks for (a) 1 µm and (b) 500 nm particle size.

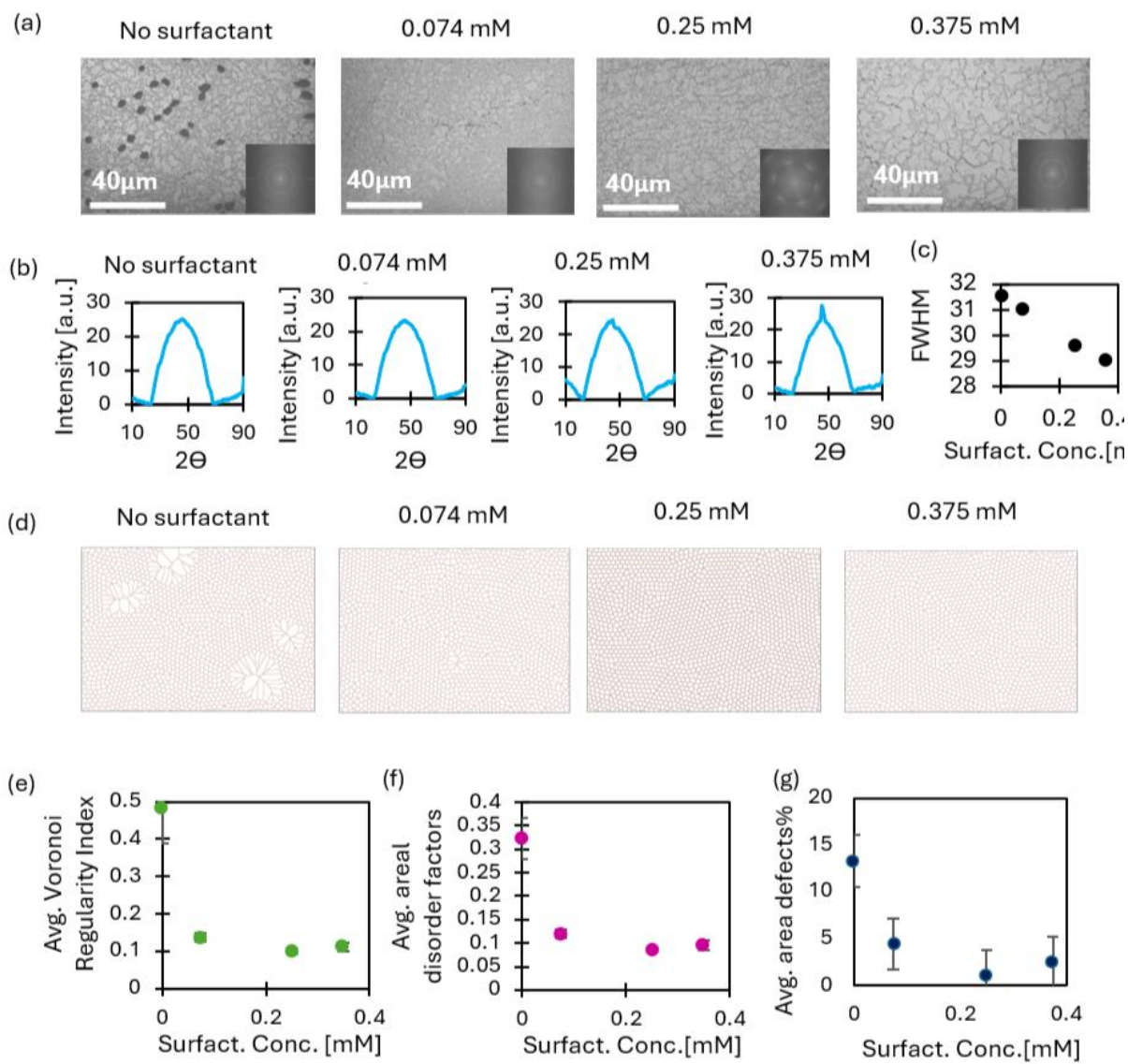

Figure S2. **Effect of Surfactant on the quality of transferred monolayer of 500 PS nanoparticles.** (a) SEM image of monolayers obtained for various surfactant concentrations. Insets show FFT patterns for each case. (b) Width of the first order FFT peaks, as a merit for the monolayer uniformity, for each surfactant concentration. (c) Width of the first order FFT peak vs. surfactant concentration. (d) Voronoi tessellations of the monolayers shown in (a). (e) Voronoi regularity index vs. surfactant concentration. (f) Areal disorder factor vs. surfactant concentration. (g) % of area occupied by defects vs. surfactant concentration. The data shown in (d)-(f) was averaged for 10 SEM images taken for each surfactant concentration.

## (a) 1 micron PS particles on PDMS

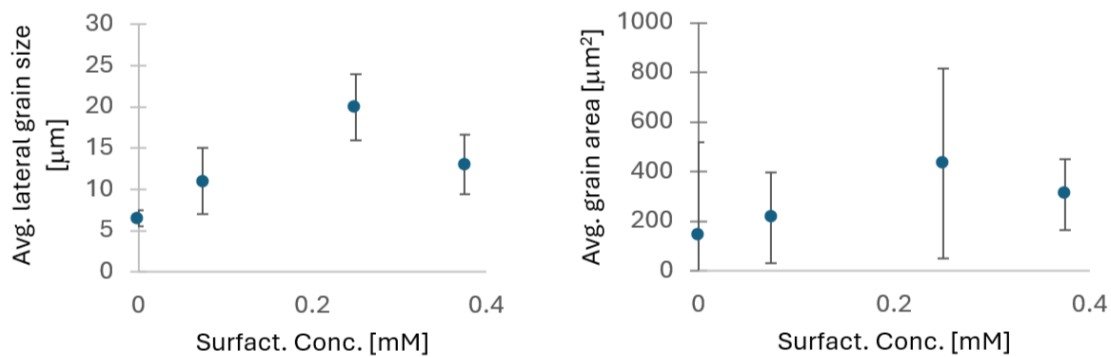

## (b) 500nm PS particles on PDMS

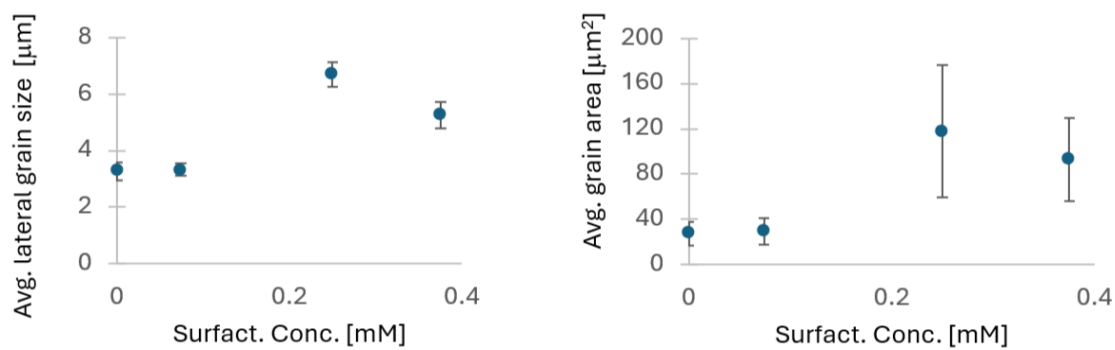

Figure S3. Average grain size and grain area of transferred monolayers of (a) 1 μm and 500 PS particles vs. surfactant concentration. The values were assessed from SEM images using ImageJ software. The lateral size of the grains was calculated based on the length of lateral lines crossing an SEM image divided by the number of grains in passes through (averaged across three lines).

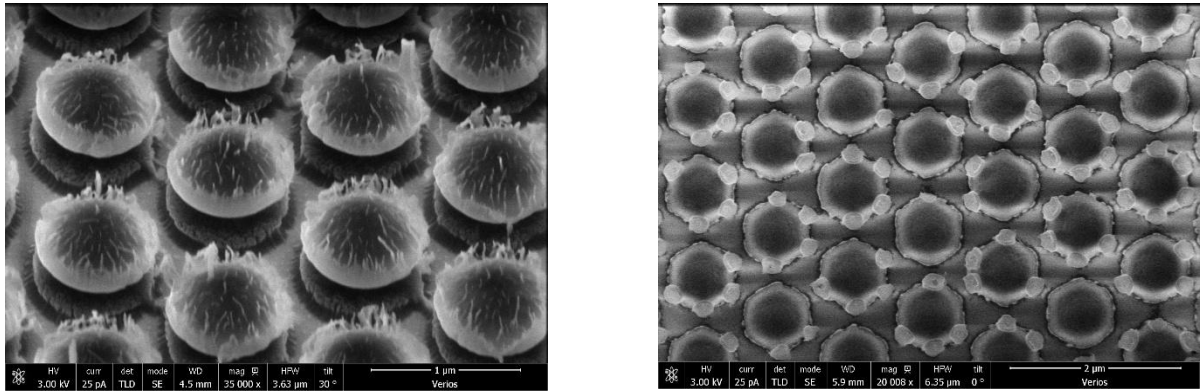

Figure S4. Examples of PEI residuals on PS spheres after pattern transfer by plasma etch of the spheres

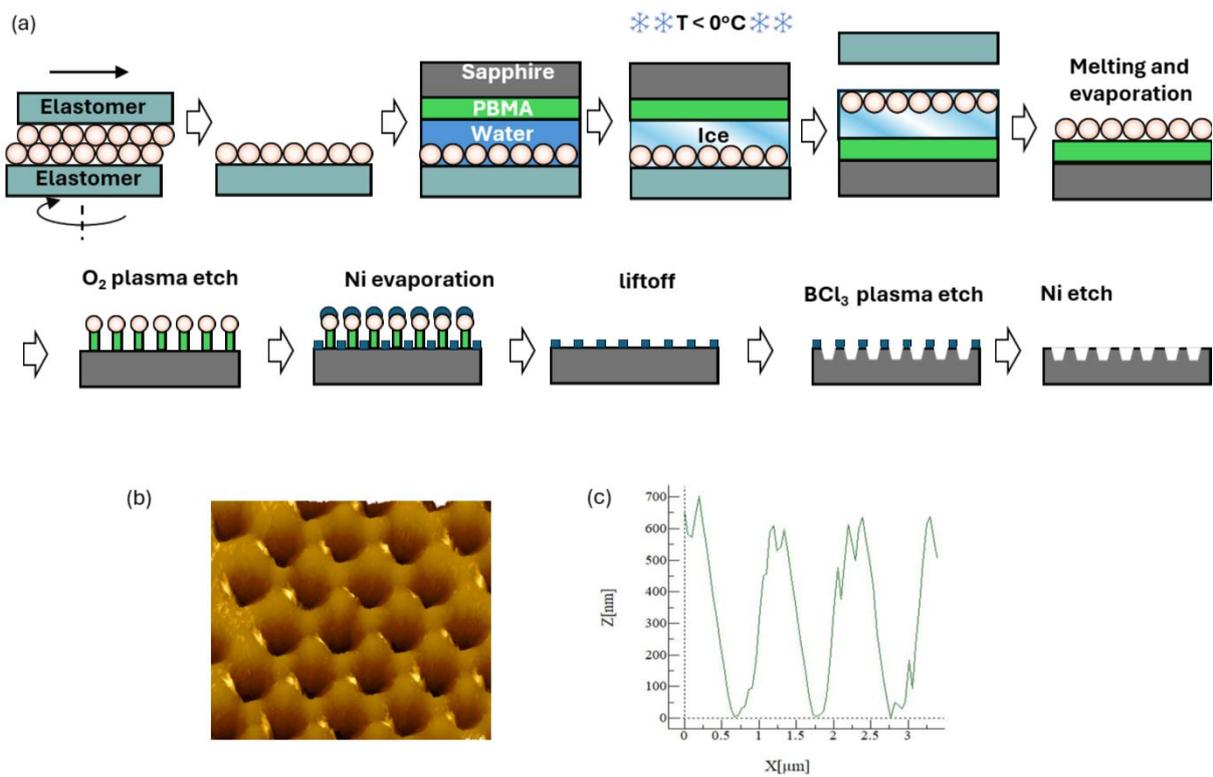

Figure. S5. (a) Fabrication process flow of moth-eye structures on sapphire using dry colloidal assembly and ice-based mechanical transfer. (b) – (c) 3D and cross section AFM of the obtained moth-eye structures on sapphire

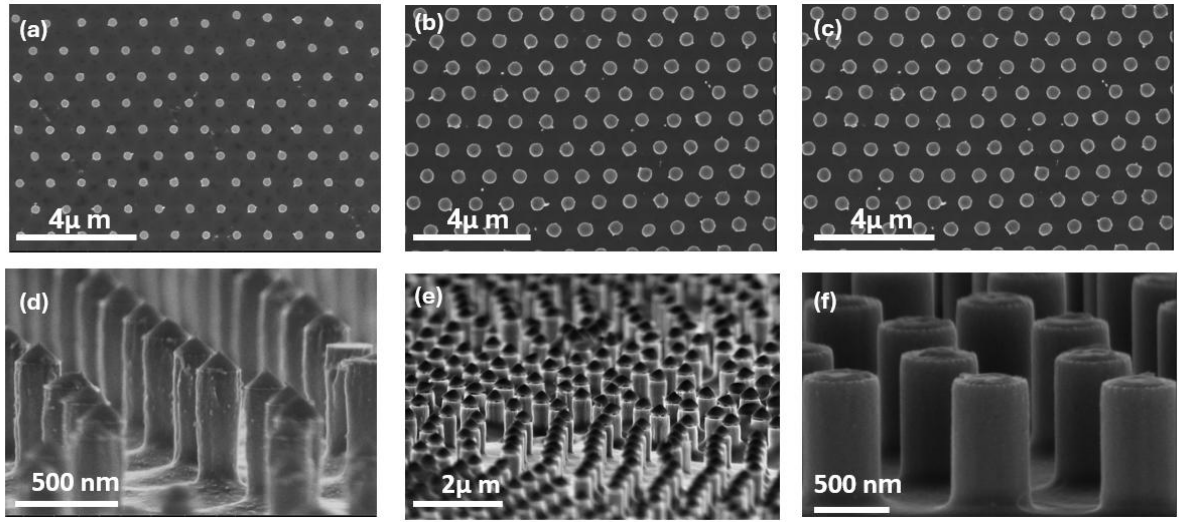

Figure S6. (a)-(c) Top view SEM of Silicon molds with used to produce arrays of pores with 200 nm, 400 nm, and 550 nm diameter in PDMS. (d)-(f) Side view of the same molds, taken at different magnifications. Note, that the Silicon pillars (d) and (e) include residuals of PS spheres on top, while in (f) PS was removed prior to SEM imaging.

## Materials Methods

### Generation of the particle monolayer

Membranes of PDMS (Sylgard 184) were prepared by mixing PDMS and its hardener in 1:10 ratio (unless otherwise specified), followed by it casting on polished silicon wafer within a petri dish, and curing at 60 °C for 1 h. The resulting PDMS was cut to pieces of desired size. A small amount (~0.3 mg per sq. inch of PDMS surface) of the nanoparticle powder was placed between two PDMS surfaces and rubbed, with the translational movement of the upper surface and rotational movement of the lower surface. A drop of water of the desired size (with or without surfactant) was placed onto the target substrate, and covered by PDMS surface containing sphere monolayer, allowing the water to spread evenly between the two surfaces. The sandwich was placed in a freezer (-18°C) for at least half an hour, then taken out to the room temperature. Immediately, PDSM was pilled off the formed ice covering the target surface, and then it left until the complete melting of ice and evaporation of water.

### Pillar based moth-eye structures on Silicon

For the antireflective structures on Silicon spectrum, the nanosphere size was first reduced from 1000 nm to 860nm by dry etching in $O_2$ plasma (Oxford Plasmapro 100 Cobra, 70 sccm $O_2$, RF=100W, P=15mTorr, t = 3min). Then, the nanosphere pattern was transferred to Silicon by dry etching using (Oxford Plasmapro 100 Estrella, RF=20 W, ICP=800 W, P=10 mTorr, 50sccm $C_4F_8$, 25sccm $SF_6$, t = 3min). Finally, the remaining microspheres were removed by sonication in chlorobenzene.

### Pore based moth-eye structures on Sapphire.

Wafers of Sapphire (2.5 cm in diameter) were first coated by PBMA using spin coating and baking, onto which an array of 1 μm polystyrene particles was transferred. Then, dry etching by O2 plasma (Oxford Plasmapro 100 Cobra, 70 sccm O2, RF = 100 W, p = 15 mTorr, t = 3:45 min) was used to reduce the particle diameter to 800 nm and create an undercut in PBMA. 270 nm nickel film was deposited by e-gun evaporation, following liftoff to obtain a Ni mask. Sapphire was etched through the mask (Oxford Plasmapro 100 Cobra, RF = 200 W, ICP = 1750 W, p = 2 mTorr, 25 sccm $BCl_3$, t = 6 min). Finally, Ni was removed by piranha solution.

**Platform for T cell activation**.

Silicon master mold was produced using 1 μm diameter polystyrene microspheres. The microsphere diameter was reduced to ~200, 400,500 nm by dry etching in O2 plasma (Oxford Plasmapro 100 Cobra, 70 sccm $O_2$, RF = 100 W, p = 20 mTorr \ Then, the microsphere pattern was transferred directly to the silicon by etching the silicon through the sphere mask (Oxford Plasmapro 100 Estrella mix gas, ICP = 800 W, RF = 20 W, 25 sccm $SF_6$, 25 sccm $C_4F_8$, p = 10 mTorr). The remaining microspheres were removed by sonication in chlorobenzene. The following steps included coating the mold with an antiadhesive agent, as described in our previous work[1]. PDMS (Sylgard 184 1:10) mixture was poured over the mold, cured for 1 hour at 60 °C, and pilled off.

**T cell activation**.

Peripheral blood mononuclear cells (PBMCs) were isolated from blood using the FICOL gradient. First, blood was diluted with PBS augmented with 2% fetal bovine serum (FBS), at a 1:1 ratio, then loaded on FICOL gradient, and centrifuged at 16 °C at 1200rpm (with no breaks or acceleration). The PBMCs were collected as the middle disc and a small portion of the underlying phase but taking care not to withdraw the pellet, washed three times with at least 1:2 with PBS 2% FBS at room temperature, and sedimented at 500 g. The cells were finally suspended in the final medium in the ratio of 2 mL per 7 mL of collected blood, counted, and diluted with the medium to final concentration of 200,000 cells per 100mL. The cells were then seeded onto ozone-treated PDMS surfaces, <2% serum and 50 units of IL-2 and left inside the incubator to adhere for 24 h.

For flow cytometry measurements, 50 000 cells were used per well. The cells were stained with respective fluorophore-conjugated antibodies at 1:1000 dilution and incubated for 30 minutes on ice. Thereafter, the cells were washed, and the dead cells were stained with DAPI (1:1000 in PBS). All the samples were analyzed on a CytoFLEX LX flow cytometer (Beckman Coulter). For analysis, the fraction of CD3-positive cells was calculated, and CD3-positive cells were then analyzed for staining with the anti CD69. The antibodies used for staining were PE anti-human CD3 and APC anti-human CD69 (all from Biolegend).